\documentclass[%
reprint,
amsmath,amssymb,
aps,longbibliography,
]{revtex4-2}

\usepackage{xcolor}
\usepackage{url}
\usepackage[colorlinks=true,linkcolor=blue,urlcolor=blue,anchorcolor=blue,citecolor=blue,bookmarksnumbered]{hyperref}
\usepackage{graphicx}
\usepackage{dcolumn}
\usepackage{bm}
\usepackage{ulem}

\begin{document}

\title{Non-Abelian Thouless pumping based on the global adiabatic criterion in Rydberg synthetic lattices}

\author{Jin-Kang~Guo$^{1}$}
\author{Jin-Lei~Wu$^{2}$}\email[]{jlwu517@zzu.edu.cn}
\author{Chuan-Cun~Shu$^{1}$}\email[]{cc.shu@csu.edu.cn}
\affiliation{$^{1}$School of Physics, Central South University, Changsha 410083, China}
\affiliation{$^{2}$Quantum Information Institute, School of Physics, Zhengzhou University, Zhengzhou 450001, China}

\begin{abstract}
We study a quantum implementation of non-Abelian Thouless pumping in Lieb lattices using Rydberg synthetic dimensions.  The lattice is encoded in twelve selected microwave-coupled Rydberg levels, forming a three-cell structure with six degenerate zero-energy states.  These zero-energy states define the working subspace for cyclic modulation of the microwave couplings, while the remaining bright states provide the dominant leakage channels at finite evolution time.  To choose the relative timing of the Gaussian pulses, we introduce a global adiabatic criterion (GAC), which evaluates the mean value and temporal fluctuation of a nonadiabatic factor obtained from a representative $\Lambda$-type transfer paradigm. With the resulting timing applied to the full twelve-level pumping dynamics, composing two elementary pumping cycles in opposite temporal orders produces distinct projected population maps. It is exactly consistent with noncommuting matrix-valued adiabatic operations in the zero-energy subspace. We numerically simulate the non-Abelian Thouless pumping using the Lindblad master equation with state-dependent Rydberg loss and representative perturbations. The results show that the GAC-selected timing within the same Gaussian pulse family gives higher target-state population than two literature-adapted Gaussian pulse schedules over the simulated parameter ranges. This quantum implementation of non‑Abelian Thouless pumping, enabled by the GAC, marks a major milestone in finite‑time geometric control and paves the way for transformative applications in holonomic quantum computing with Rydberg synthetic lattices.
\end{abstract}
\maketitle

\section{Introduction}
\label{sec:introduction}

Thouless pumping provides a standard route for converting cyclic parameter modulation into geometric transport~\cite{Thouless1983Pumping,Kraus2012PRL,Zilberberg2018Nature,Lohse2018Nature,Jurgensen2021Nature,Walter2023HubbardThoulessBreakdown,Citro2023NRP,ZZhu2024Science,ZCheng2025NC,QMo2025PRL}.  For an isolated adiabatic band, the transported quantity is governed by Abelian geometric phases and, in periodic systems, by the associated Berry curvature or Chern number~\cite{Berry1984Phase,Simon1983Holonomy}. A richer situation arises when the evolution takes place inside a degenerate subspace. In that case the adiabatic operation is represented by a matrix acting on the degenerate states, i.e., the Wilczek-Zee holonomy~\cite{Wilczek1984Holonomy,Leroux2018NonAbelianAMO,Unanyan1999NonAbelianAdiabatic,Rigolin2014DegenerateAdiabatic,Wu2021HolonomicEntanglingGates,Yun2024SiVGeometricGates}.  Because such matrix-valued operations need not commute, reversing the order of two pumping cycles can change the projected output state. This order dependence underlies non-Abelian Thouless pumping~\cite{Brosco2021PRA,Danieli2025AVSQS}. Its implementation has been explored in classical wave systems using photonic and acoustic waveguides~\cite{Sun2022NATP,You2022NonAbelianPump,YKSun2024NC,YChen2025NC,ZLShan2025SB}. However, a feasible scheme for implementing the non-Abelian Thouless pumping in quantum systems remains unexplored.

Rydberg synthetic dimensions provide a controllable way to encode lattice sites in selected atomic levels and to use microwave transitions as programmable hopping channels~\cite{Ostmann2018RydbergFlatBand,Ozawa2019NRP,DYLi2025PI}.  The amplitudes, phases, and detunings of these transitions can be tuned independently, making it possible to engineer synthetic tight-binding lattices with designed connectivity and gauge structure~\cite{Scholl2022RydbergXXZ,Wu2022RydbergGaugeFlux,Wu2025FloquetRydbergInteractions,Wu2020RydbergAmplitudeModulation,Zhou2026PTSSHQuantumBattery}.  Such control has enabled Rydberg implementations of topological insulators, quantum walks, synthetic gauge fluxes, flat-band geometries, and Aharonov-Bohm caging~\cite{Kanungo2022RydbergSyntheticDimension,YLu2024PRA,TChen2024PRL,YLu2024PRA2,Chen2024RydbergFlux,Chen2025RydbergCage}. Thouless pumping has also been explored in a synthetic Rydberg dimension~\cite{Trautmann2024RydbergThouless,huang2025interactionassistedtopologicalpumpingfew}. These results motivate using microwave-coupled Rydberg levels as a finite synthetic lattice rather than only as a multilevel spectroscopic system.

The same level-resolved control also introduces finite-time constraints.  A pumping cycle should remain inside the degenerate zero-energy subspace, whereas a finite modulation time may drive leakage to the bright states.  Making the evolution sufficiently slow is not favorable for Rydberg levels, where radiative decay, imperfect state preparation, off-resonant microwave couplings, and residual detunings synergistically reduce the final target-state population~\cite{Beterov2009RydbergLifetime,Leseleuc2018RydbergImperfections}.  At the same time, increasing all microwave amplitudes is limited by spectral selectivity among nearby Rydberg transitions.  The control problem is therefore not only to specify a cyclic path, but also to choose its timing so that adiabatic following is retained without unnecessary exposure to loss~\cite{Fan2023MolecularPolaritonControl,Hong2025MolecularRotationControl}.

In this work, we propose to implement the non-Abelian Thouless pumping in a synthetic Lieb lattice using internal quantum states of a Rydberg atom.
We use microwave-coupled Rydberg levels to construct a finite three-cell synthetic Lieb lattice and study finite-time non-Abelian  Thouless pumping in its degenerate zero-energy subspace.  The six zero-energy states form the working basis for cyclic microwave modulation, whereas the bright states determine the main leakage channels during finite-time evolution. To address the finite-time constraint, we introduce a global adiabatic criterion (GAC) for the local adiabatic-passage steps~\cite{Xiao2026GACPhotonicPumping,wu2026globaladiabaticcriterionfast}.  It evaluates the nonadiabatic factor over the full transfer window and uses its mean value and temporal fluctuation to guide the timing choice.  Unlike shortcut-to-adiabaticity protocols that introduce additional counterdiabatic or auxiliary couplings~\cite{Demirplak2003JPCA,Chen2010ShortcutAdiabaticPassage,GueryOdelin2019STA,Guo2024OptomechanicalSTA}, this approach can keep the Gaussian microwave-coupling envelopes fixed and selects only their relative timing. Two elementary Gaussian pumping cycles are composed in opposite temporal orders to probe the order-dependent projected dynamics generated within the zero-energy subspace.  Their relative pulse timing is selected by the GAC through local $\Lambda$-type transfer steps and is then used in the full twelve-level model.  Closed-system dynamics, open-system simulations, and perturbation scans compare this schedule with two literature-adapted Gaussian pulse schedules~\cite{Longhi2019PRB,Tian2022PRL}, validating more favorable finite-time target-state population and robustness of the GAC-selected schedule.

The paper proceeds from the Rydberg Lieb construction in Sec.~\ref{sec:model} to the elementary cycles and GAC-based timing design in Sec.~\ref{sec:single_cycle}, the composed-cycle response in Sec.~\ref{sec:nonabelian_composition}, and open-system robustness in Sec.~\ref{sec:robustness}; appendices give derivations and numerical definitions.

\section{Rydberg Lieb lattice and zero-energy subspace}
\label{sec:model}

\begin{figure}[t]
    \centering
    \includegraphics[width=0.95\linewidth]{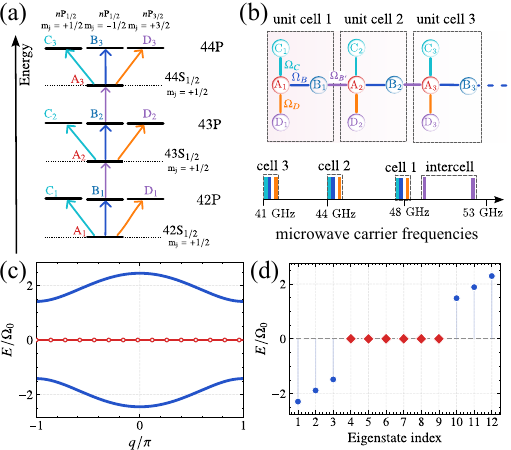}
    \caption{Rydberg level construction forming a finite synthetic Lieb lattice.
    (a) Selected ${}^{39}\mathrm{K}$ Rydberg levels used to encode the four states in each unit cell, with three neighboring principal quantum numbers forming the finite lattice.
    (b) Effective three-cell Lieb connectivity generated by microwave-induced couplings, together with the corresponding microwave carrier-frequency groups used to address cell-resolved and intercell transitions.
    (c) Bulk spectrum of the corresponding periodic Lieb lattice, showing two degenerate zero-energy flat bands separated from the bright branches.
    (d) Spectrum of the finite three-cell lattice, where the six zero-energy states form the working subspace used for the pumping protocols.
    }
    \label{fig:fig1}
\end{figure}

We first specify a concrete implementation for the finite synthetic Lieb lattice using ${}^{39}\mathrm{K}$ Rydberg levels~\cite{TChen2024PRL,Chen2024RydbergFlux,Chen2025RydbergCage}.  As shown in Fig.~\ref{fig:fig1}(a), the three unit cells are encoded in the principal-quantum-number groups $n_i=41+i$, with $i=1,2,3$.  The four sites in the $i$-th unit cell are denoted by $|A_i\rangle$, $|B_i\rangle$, $|C_i\rangle$, and $|D_i\rangle$, corresponding to $|n_iS_{1/2},m_j=+1/2\rangle$, $|n_iP_{1/2},m_j=+1/2\rangle$, $|n_iP_{1/2},m_j=-1/2\rangle$, and $|n_iP_{3/2},m_j=+3/2\rangle$, respectively.  Thus $N=3$ gives twelve selected Rydberg states.  The microwave carrier frequencies select the intended transitions, while the microwave Rabi amplitudes set the effective hopping matrix elements.  The frequency groups indicated in Fig.~\ref{fig:fig1}(b) separate the cell-resolved transitions and the intercell channel.  While the concrete level set considered here is based on ${}^{39}\mathrm{K}$, the required level-resolved microwave control is consistent with recent cold-atom Rydberg synthetic-dimension experiments, including synthetic gauge fluxes, flat-band interference, Aharonov--Bohm caging, and Thouless pumping~\cite{Kanungo2022RydbergSyntheticDimension,Chen2024RydbergFlux,Chen2025RydbergCage,Trautmann2024RydbergThouless}.

The target connectivity is the finite Lieb lattice shown in Fig.~\ref{fig:fig1}(b).  Within each unit cell, $|A_i\rangle$ is coupled to $|B_i\rangle$, $|C_i\rangle$, and $|D_i\rangle$ through hopping rates $\Omega_B$, $\Omega_C$, and $\Omega_D$, respectively.  Neighboring cells are connected by the intercell hopping $\Omega_{B'}$ between $|B_i\rangle$ and $|A_{i+1}\rangle$.  In the rotating frame and under resonant driving, the rotating-wave approximation gives the finite tight-binding Lieb-lattice Hamiltonian~(see derivation in Appendix~\ref{app:finite_lieb_hamiltonian})
\begin{equation}
\begin{aligned}
\hat{H}(t)
=&
\sum_{i=1}^{N}
\sum_{\mu=B,C,D}
\Omega_{\mu}(t)
|A_i\rangle\langle \mu_i|
\\
&+
\sum_{i=1}^{N-1}
\Omega_{B'}(t)
|A_{i+1}\rangle\langle B_i|
+
\mathrm{H.c.}
\end{aligned}
\label{eq:finite_lieb_hamiltonian}
\end{equation}
The selected levels also indicate how the couplings can be addressed in experiment.  The microwave channels can be selected by carrier frequency, polarization selection rules, and a weak bias field that defines the quantization axis.  Although $|B_i\rangle$ and $|C_i\rangle$ belong to the same $n_iP_{1/2}$ fine-structure level in the zero-field limit, their couplings can be distinguished by polarization, while $|D_i\rangle$ is separated by the $n_iP_{3/2}$ fine-structure splitting.

The zero-energy sector follows directly from the way the selected Rydberg levels are connected by the microwave fields.  In each unit cell, the state $|A_i\rangle$ is coupled only to the three branch states $|B_i\rangle$, $|C_i\rangle$, and $|D_i\rangle$, while the intercell channel couples $|B_i\rangle$ to $|A_{i+1}\rangle$.  Thus, after grouping the basis into the $A$-sublattice and the $(B,C,D)$-sublattice, Eq.~\eqref{eq:finite_lieb_hamiltonian} has an off-diagonal bipartite form, $H=\begin{pmatrix}0&K\\K^\dagger&0\end{pmatrix}$, with $K$ being an $N\times3N$ coupling matrix.  For generic nonzero couplings in this Lieb geometry, the sublattice imbalance leaves $2N$ zero-energy states, as detailed in Appendix~\ref{app:zero_mode_counting}.  The periodic reference spectrum in Fig.~\ref{fig:fig1}(c) shows the corresponding flat-band origin of this zero-energy sector, while the finite three-cell spectrum in Fig.~\ref{fig:fig1}(d) shows the six zero-energy states.

These six zero-energy states form the working subspace for the pumping dynamics.  Under a cyclic modulation of the microwave-induced couplings, adiabatic following confines the evolution to this degenerate subspace, where the geometric operation is matrix-valued rather than a scalar Berry phase~\cite{Brosco2021PRA,Danieli2025AVSQS,Sun2022NATP,You2022NonAbelianPump,YKSun2024NC,YChen2025NC,ZLShan2025SB}.  Reversing the order of two cyclic operations can therefore change the projected final state, leading to the order-dependent response.  At finite modulation time, the main error channel is leakage from the zero-energy subspace to the bright states separated in Fig.~\ref{fig:fig1}(d), which motivates the GAC-based timing design.
\section{Elementary pumping cycles and GAC-based timing design}
\label{sec:single_cycle}

\begin{figure}[t]
    \centering
    \includegraphics[width=0.95\linewidth]{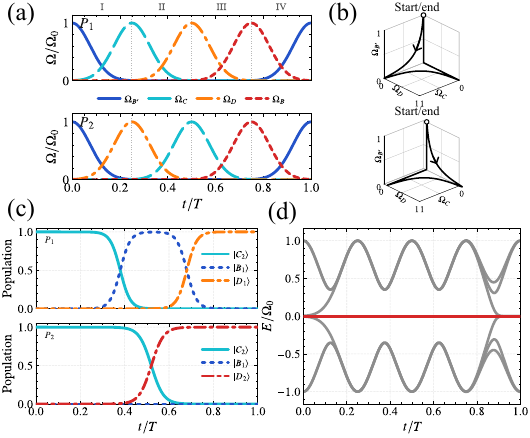}
    \caption{
    Elementary single-cycle pumping protocols.
    (a) Gaussian microwave-coupling sequences for the two cycles $P_1$ and $P_2$.  Both cycles use the same four coupling channels $\Omega_B$, $\Omega_C$, $\Omega_D$, and $\Omega_{B'}$; $P_2$ is obtained from $P_1$ by exchanging the temporal order of the $\Omega_C$ and $\Omega_D$ pulses.
    (b) Corresponding closed trajectories in the projected $(\Omega_C,\Omega_D,\Omega_{B'})$ parameter space.  The two cycles have the same start and end point but follow different intermediate paths.
    (c) Representative closed-system population dynamics for the same initial state $|C_2\rangle$.  In $P_1$, the population is transferred mainly to $|D_1\rangle$ through intermediate occupation of $|B_1\rangle$, whereas in $P_2$ it is transferred mainly to $|D_2\rangle$.
    (d) Representative instantaneous spectrum during the cycle.  The zero-energy states remain separated from the bright branches, defining the gap relevant for adiabatic following.
    }
    \label{fig:fig2}
\end{figure}

The finite zero-energy subspace established above provides the basis for defining the elementary pumping cycles used throughout this work.  A pumping cycle is a closed trajectory of the microwave-coupling vector
$\boldsymbol{\Omega}(t)=(\Omega_{B'}(t),\Omega_C(t),\Omega_D(t),\Omega_B(t))$
in the coupling-parameter space.  Figure~\ref{fig:fig2}(a) shows two such Gaussian cycles, denoted by $P_1$ and $P_2$.  They use the same four coupling channels, but $P_2$ is obtained from $P_1$ by exchanging the temporal order of the $\Omega_C$ and $\Omega_D$ pulses.  The corresponding closed paths in the projected parameter space are shown in Fig.~\ref{fig:fig2}(b).
We describe the induced operation in the zero-energy sector spanned by the ordered states $|C_1\rangle, |D_1\rangle,\ldots, |C_N\rangle, |D_N\rangle$.  An initial state in this sector can be written as $|\psi_0(0)\rangle=\sum_{i=1}^{N}(c_i|C_i\rangle+d_i|D_i\rangle)$.  In the adiabatic limit, one cycle of $P_\nu$ maps this state to $|\psi_0(T_\nu)\rangle=e^{{\rm i}\chi_\nu}\hat U_\nu|\psi_0(0)\rangle$ with $\nu=1,2$, where $\hat U_\nu$ is the Wilczek-Zee operation generated within the zero-energy subspace and $\chi_\nu$ is an overall phase.  Thus $\hat U_2\hat U_1$ denotes the composed operation obtained by applying $P_1$ first and $P_2$ second, while $\hat U_1\hat U_2$ denotes the reversed temporal order. Figure~\ref{fig:fig2}(c) gives representative closed-system population traces for the initial state $|C_2\rangle$.  Since $P_1$ and $P_2$ realize different projected operations, $\hat U_1|C_2\rangle$ and $\hat U_2|C_2\rangle$ lead to different final components.  In the full dynamics, $P_1$ transports population mainly to $|D_1\rangle$ with transient $|B_1\rangle$ occupation, whereas $P_2$ transports it mainly to $|D_2\rangle$.
The instantaneous spectrum in Fig.~\ref{fig:fig2}(d) shows the gap separating the degenerate manifold from the bright branches during the cycle.  Because $P_1$ and $P_2$ use the same envelopes and differ only by the $\Omega_C$-$\Omega_D$ ordering, one spectrum suffices for both.  The finite-time errors considered below are therefore mainly leakage across this gap.

\begin{figure}[t]
    \centering
    \includegraphics[width=0.95\linewidth]{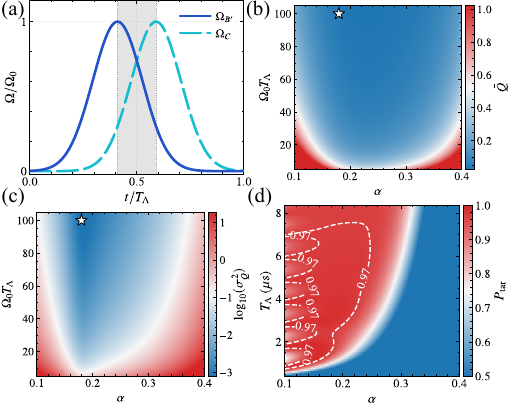}
    \caption{
    GAC-guided timing selection for local Gaussian microwave couplings.
    (a) Gaussian coupling envelopes $\Omega_{B'}$ and $\Omega_C$ for the representative local adiabatic-passage step.  The shaded region indicates the pulse overlap controlled by the delay parameter $\alpha$.
    (b) Mean nonadiabaticity $\overline{Q}$ as a function of $\alpha$ and the dimensionless local duration $\Omega_0T_\Lambda$.
    (c) Temporal fluctuation of the nonadiabaticity, shown as $\log_{10}(\sigma_Q^2)$, over the same parameter space.
    (d) Target-state population $P_{\mathrm{tar}}$ obtained from the Lindblad master-equation calculation for the same local step, with input state $|C_2\rangle$ and target state $|B_1\rangle$, plotted as a function of $\alpha$ and the local duration $T_\Lambda$.
    }
    \label{fig:fig3}
\end{figure}

For a fixed input state, each active part of the pumping cycle in the whole Lieb lattice can be reduced to a local $\Lambda$-type adiabatic-passage step, as in flat-band Lieb pumping protocols~\cite{Sun2022NATP}.  The local Hamiltonian, zero-energy state, mixing angle, and nonadiabaticity factor $Q(t)$ for the GAC theory are given in Appendix~\ref{app:local_three_level_reduction}.  Here the reduction identifies the transfer window and leakage mechanism. The population traces and composed-cycle calculations use the complete twelve-level Hamiltonian.  Figure~\ref{fig:fig3} applies the GAC theory to the local adiabatic-passage step.  As shown in Fig.~\ref{fig:fig3}(a), the target-side coupling $\Omega_{B'}(t)$ precedes the input-side coupling $\Omega_C(t)$, giving the counterintuitive ordering of a $\Lambda$-type adiabatic passage~\cite{Bergmann1998STIRAP}.  The couplings are taken as 
\begin{eqnarray}
\Omega_{B'}(t)=\Omega_0\exp[-(t-t_{B'})^2/\tau^2],\nonumber\\
\Omega_C(t)=\Omega_0\exp[-(t-t_C)^2/\tau^2],
\end{eqnarray}
 where $t_{B'}=(1-\alpha)T_\Lambda/2$, $t_C=(1+\alpha)T_\Lambda/2$, and $\tau=(1-\alpha)T_\Lambda/5$.  Thus $\alpha$ sets the delay and overlap within this fixed Gaussian family.

The local model isolates the finite-time leakage channel used in the GAC theory.  In the adiabatic limit, the local zero-energy state connects $|C_2\rangle$ and $-|B_1\rangle$ while avoiding $|A_2\rangle$.  At finite duration, however, the changing mixing angle couples it to local bright states.  The nonadiabatic factor $Q(t)$, defined in Eq.~\eqref{eq:Q_non} of Appendix~\ref{app:local_three_level_reduction}, characterizes the finite-time leakage from the instantaneous zero-energy state to the bright states.  The GAC theory uses the global distribution of $Q(t)$ over the full transfer window, so we define
\begin{equation}
\begin{aligned}
\overline Q
&=
\frac{1}{T_\Lambda}
\int_0^{T_\Lambda}
Q(t)\,\mathrm{d}t,
\\
\sigma_Q^2
&=
\frac{1}{T_\Lambda}
\int_0^{T_\Lambda}
\left[
Q(t)-\overline Q
\right]^2
\,\mathrm{d}t .
\end{aligned}
\label{eq:Q_mean_variance}
\end{equation}
For the local transfer considered here, the modulation duration is identified with $T_\Lambda$.
Here $\overline Q$ measures the average nonadiabatic tendency, whereas the variance $\sigma_Q^2$ quantifies the temporal fluctuation or concentration of this tendency over the transfer window.  As summarized in Appendix~\ref{app:gac_theory}, the GAC theory gives a conservative design estimate for the total nonadiabatic transition probability
\begin{equation}
\varepsilon
\le
\left[
\Delta E_{\min}T_\Lambda
\sqrt{\overline Q^2+\sigma_Q^2}
\right]^2 ,
\end{equation}
where $\varepsilon$ is the total nonadiabatic transition probability and $\Delta E_{\min}$ is the minimum gap between the local zero-energy state and the bright states within the transfer window.  In a fixed Gaussian pulse family, lowering $\sigma_Q^2$ suppresses localized nonadiabatic leakage peaks and supports faster finite-time adiabatic transfer.
Figures~\ref{fig:fig3}(b) and \ref{fig:fig3}(c) show the GAC landscapes of $\overline Q$ and $\log_{10}(\sigma_Q^2)$ as functions of $\alpha$ and the dimensionless local duration $\Omega_0T_\Lambda$.  The selected value $\alpha=0.18$ reduces temporal fluctuations of the nonadiabaticity factor without leaving from the low-$\overline Q$ region. Figure~\ref{fig:fig3}(d) checks the same local step for input $|C_2\rangle$ by plotting the population $P_{\mathrm{tar}}$ of the target state $|B_1\rangle$ versus $\alpha$ and the local duration $T_\Lambda$. The data are obtained by numerically solving the Lindblad master equation detailed in Appendix~\ref{app:open_system_master_equation}, where the transition frequencies, dipole matrix elements, and lifetimes used in the numerical estimates are calculated by Alkali Rydberg Calculator (ARC)~\cite{Sibalic2017ARC}.  The high-$P_{\mathrm{tar}}$ region overlaps with the selected timing window around $\alpha=0.18$, supporting the corresponding GAC-selected schedule in the complete twelve-level pumping calculations.

\begin{figure}[t]
	\centering
	\includegraphics[width=0.95\linewidth]{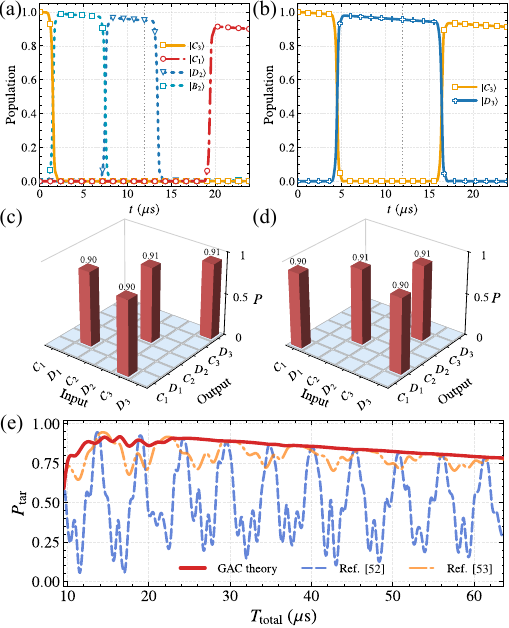}
	\caption{Order-dependent response of composed pumping cycles with Lindblad master-equation simulations. Time-resolved populations for the temporal orders (a)~$\hat U_2\hat U_1$ and (b)~$\hat U_1\hat U_2$, starting from the same input state $|C_3\rangle$.
	Projected transfer maps for (c)~$\hat U_2\hat U_1$ and (d)~$\hat U_1\hat U_2$, respectively, in the computational basis $\{|C_1\rangle,|D_1\rangle,|C_2\rangle,|D_2\rangle,|C_3\rangle,|D_3\rangle\}$.
		The bar height gives the final projected population for each input-output pair.
		(e) Target-state population $P_{\mathrm{tar}}$ as a function of the full two-cycle duration $T_{\mathrm{total}}$, averaged over the two temporal orders and the two input states $|C_3\rangle$ and $|D_3\rangle$, with the target state chosen according to the corresponding composed-cycle mapping. 
		The GAC-selected schedule is compared with two reference Gaussian pulse schedules based on Refs.~\cite{Longhi2019PRB,Tian2022PRL}.}
	\label{fig:fig4}
\end{figure}
\section{Order-dependent non-Abelian Thouless pumping}
\label{sec:nonabelian_composition}

To examine the order dependence generated in the zero-energy subspace, the two elementary pumping cycles are composed in opposite temporal orders.  The single-cycle operations are $\hat U_1$ and $\hat U_2$.  With the usual convention that the rightmost operation acts first, $\hat U_2\hat U_1$ corresponds to applying $P_1$ followed by $P_2$, while $\hat U_1\hat U_2$ corresponds to the reversed temporal order.  The two composed protocols use the same microwave-coupling channels and elementary cycles.  The composed-cycle results in Fig.~\ref{fig:fig4} are obtained from Lindblad master-equation simulations with state-dependent Rydberg decay.  Figures~\ref{fig:fig4}(a) and \ref{fig:fig4}(b) show the resulting time-resolved populations for the same input state $|C_3\rangle$.  For $\hat U_2\hat U_1$, the population is transported across the finite lattice and ends predominantly in $|C_1\rangle$, with transient occupation of intermediate zero-energy components.  For $\hat U_1\hat U_2$, it first moves mainly into $|D_3\rangle$ but then returns to $|C_3\rangle$.  These different outputs arise from reversing the order of the two cyclic operations, rather than from changing the available couplings.  The projected transfer maps in Figs.~\ref{fig:fig4}(c) and \ref{fig:fig4}(d) extend this comparison to the six-dimensional computational basis
$\{|C_1\rangle,|D_1\rangle,|C_2\rangle,|D_2\rangle,|C_3\rangle,|D_3\rangle\}$.
Each column corresponds to an input basis state, and each bar gives the final projected population in an output basis state.  The dominant bars form different transfer patterns for $\hat U_2\hat U_1$ and $\hat U_1\hat U_2$, showing that the two compositions act fully differently on the zero-energy subspace.  The order-dependent maps provide a population-level signature consistent with noncommuting matrix-valued adiabatic operations.

Figure~\ref{fig:fig4}(e) compares the finite-time performance of the composed protocol. We estimate $P_{\mathrm{tar}}$ versus the full two-cycle duration $T_{\mathrm{total}}$, averaged over the four cases formed by the two temporal orders and the two input states $|C_3\rangle$ and $|D_3\rangle$, with the target state chosen from the corresponding composed-cycle mapping.  All schedules use the same Gaussian pulse family: the reference schedules are adapted from earlier adiabatic-passage pumping and dynamically modulated transfer protocols~\cite{Longhi2019PRB,Tian2022PRL}, whereas the GAC-selected schedule uses the global distribution of the local nonadiabatic factor through $\overline Q$ and $\sigma_Q^2$.  The GAC-selected target-state population curve remains comparatively smooth and high over the plotted physical time window, whereas the reference schedules show stronger finite-time oscillations or lower target population. Thus Fig.~\ref{fig:fig4} links the order-dependent projected maps with the open-system duration scan: the GAC-selected schedule gives higher finite-time target-state population for the same composed cycles, and the differing maps are consistent with noncommuting adiabatic operations in the zero-energy subspace.

\section{Robustness}
\label{sec:robustness}

\begin{figure}[t]
    \centering
    \includegraphics[width=0.95\linewidth]{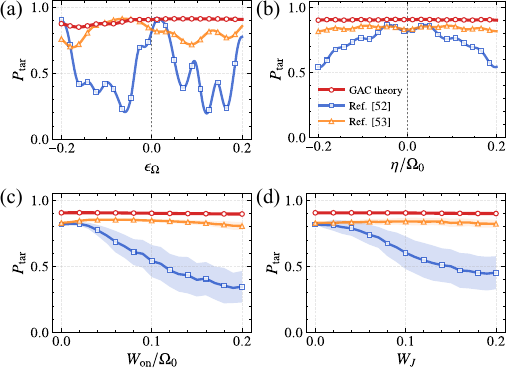}
    \caption{
    Open-system response of the composed pumping protocol to representative imperfections.
    (a) Global Rabi-amplitude error $\epsilon_\Omega$.
    (b) Deterministic detuning gradient $\eta/\Omega_0$.
    (c) Zero-mean static random on-site disorder $W_{\mathrm{on}}/\Omega_0$.
    (d) Static relative coupling disorder $W_J$.
    In (c,d), the shaded regions denote $\pm 1$ standard deviation over $N_{\mathrm{real}}=200$ disorder realizations. More details are given in Appendix~\ref{app:open_system_master_equation}.
    }
    \label{fig:fig5}
\end{figure}

The robustness of the full composed pumping protocol is examined in the presence of Rydberg state decay and representative perturbations.  The decay-included dynamics is simulated using the Lindblad master equation with state-dependent Rydberg decay rates estimated at $10~\mathrm{K}$.  The Hilbert space consists of the twelve selected Rydberg working states and an auxiliary sink state that collects population lost from the selected Rydberg states. The model details are given in Appendix~\ref{app:open_system_master_equation}.  The target-state population $P_{\mathrm{tar}}$ is averaged over the two input states $|C_3\rangle$, $|D_3\rangle$ and over the two temporal orders.  Since $P_{\mathrm{tar}}$ is not normalized by the survival probability, it includes both coherent transfer errors and population loss from the selected Rydberg levels.

All curves in Fig.~\ref{fig:fig5} are evaluated at the same total duration, $T_{\mathrm{total}}\simeq23.9\,\mu\mathrm{s}$, corresponding to $\Omega_0T_{\mathrm{total}}=300$.  The two reference curves use literature-adapted Gaussian pulse schedules~\cite{Longhi2019PRB,Tian2022PRL}.  Figures~\ref{fig:fig5}(a) and \ref{fig:fig5}(b) show the response to systematic imperfections.  A global Rabi-amplitude error changes the overall coupling scale, while a deterministic detuning gradient adds coherent diagonal offsets across the synthetic lattice.  Over the plotted ranges, the GAC-selected schedule retains a higher target-state population and weaker finite-time oscillations than the reference schedules.
Figures~\ref{fig:fig5}(c) and \ref{fig:fig5}(d) test static random perturbations.  The two panels distinguish diagonal disorder of the selected Rydberg levels from relative disorder in the microwave-induced couplings.  In both scans, the GAC-selected schedule shows slower degradation and a narrower disorder-induced spread, while the reference schedules show either broader spread or lower target-state population.
Across the tested perturbation ranges, the GAC-selected schedule provides a robustness advantage for the full composed protocol.  Since all three schedules use the same pumping geometry and the same total duration, the observed differences reflect the relative timing of the Gaussian pulses rather than a change of the target operation.  The higher $P_{\mathrm{tar}}$ and weaker fluctuations of the GAC-selected schedule are consistent with reduced accumulated leakage to bright states during the full twelve-level sequence.

\section{Conclusion}
We have studied the quantum implementation of finite-time non-Abelian pumping in a Lieb lattice constructed from the synthetic dimension using selected microwave-coupled Rydberg levels.  The finite three-cell lattice supports six zero-energy states, which form the degenerate working subspace for the cyclic microwave modulation.  In this subspace, adiabatic evolution is described by non-Abelian matrix-valued operations, allowing the temporal order of elementary pumping cycles to affect the projected final state. The timing of the Gaussian microwave couplings was chosen using the GAC theory.  By reducing the active segments of the Lieb pumping cycle to local $\Lambda$-type adiabatic-passage steps, the criterion relates finite-time leakage to the mean value and temporal fluctuation of the local nonadiabatic factor.  This provides a timing-selection rule within the same Gaussian pulse family, rather than introducing additional counterdiabatic couplings or auxiliary controls.
The composed-cycle simulations show distinct projected population responses for the two temporal orders, consistent with noncommuting adiabatic operations in the zero-energy subspace.  In the open-system calculations, the GAC-selected schedule maintains higher target-state population than the two literature-adapted Gaussian pulse schedules under state-dependent Rydberg decay, systematic calibration errors, detuning gradients, and static disorder.  This improvement is obtained without changing the pumping geometry or the pulse family, indicating that the relative pulse timing selected by the GAC is the key element for suppressing finite-time bright-state leakage in the complete composed sequence.  These results support Rydberg synthetic dimensions as a controllable setting for finite-time coherent operations in degenerate working subspaces, with possible extensions to microwave-selectivity effects, bias-field-induced shifts, interacting Rydberg systems, and larger synthetic lattices.

\begin{acknowledgments}
    The authors acknowledge financial support from the  National Natural Science Foundations of China (NSFC) under Grants No. 62571494, No. 12304407, and No. 12274470, from the Natural Science Foundation of Henan Province under Grant No. 262300421244, and from the
China Postdoctoral Science Foundation under Grant No. 2024M762973.
\end{acknowledgments}
\appendix
\section{Finite synthetic Lieb model and zero-energy sector}
\label{app:finite_lieb_hamiltonian}
\label{app:periodic_lieb_bands}
\label{app:zero_mode_counting}

This appendix connects the microwave-coupled Rydberg levels to the finite synthetic Lieb Hamiltonian used in the main text.  For selected Rydberg states $|r\rangle$ and $|r'\rangle$, we define the transition operator
$\hat\sigma_{r,r'}\equiv |r\rangle\langle r'|$.  In the laboratory frame, the Hamiltonian of the selected levels driven by microwave fields can be written as
\begin{equation}
\begin{aligned}
\hat H_{\rm S}(t)
=&
\sum_r \omega_r \hat\sigma_{r,r}
\\
&+
\sum_{\langle r,r'\rangle_{\rm mw}}
\left[
\Omega_{r,r'}(t)
e^{{\rm i}\omega_{a,rr'}t}
\hat\sigma_{r,r'}
+
{\rm H.c.}
\right],
\end{aligned}
\label{eq:rydberg_lab_hamiltonian}
\end{equation}
where $\omega_r$ is the energy of the selected Rydberg level $|r\rangle$, $\omega_{a,rr'}$ is the microwave carrier frequency, and $\langle r,r'\rangle_{\rm mw}$ denotes the set of driven transitions.  The complex Rabi amplitude $\Omega_{r,r'}(t)$ contains the microwave envelope and phase.

Transforming Eq.~\eqref{eq:rydberg_lab_hamiltonian} into a near-resonant rotating frame and applying the rotating-wave approximation gives the interaction-picture Hamiltonian
\begin{equation}
\hat H_{\rm I}(t)
=
\sum_r \Delta_r \hat\sigma_{r,r}
+
\sum_{\langle r,r'\rangle_{\rm mw}}
\left[
\Omega_{r,r'}(t)
\hat\sigma_{r,r'}
+
{\rm H.c.}
\right].
\label{eq:rydberg_rotating_hamiltonian}
\end{equation}
Here $\Delta_r$ denotes the residual detuning of state $|r\rangle$ in the chosen rotating frame.  These detunings act as effective on-site potentials in the synthetic lattice.  In the resonant pumping model used in the main text, the ideal detunings are set to zero, while controlled or residual detunings are included separately in the perturbation analysis.

The finite Lieb lattice is obtained by choosing the driven transition graph $\langle r,r'\rangle_{\rm mw}$ according to the connectivity in Fig.~\ref{fig:fig1}(b).  Each unit cell contains four selected Rydberg states,
$|A_i\rangle$, $|B_i\rangle$, $|C_i\rangle$, and $|D_i\rangle$.  The intracell microwave fields couple $|A_i\rangle$ to $|B_i\rangle$, $|C_i\rangle$, and $|D_i\rangle$, with effective hopping amplitudes $\Omega_B(t)$, $\Omega_C(t)$, and $\Omega_D(t)$, respectively.  The intercell microwave field couples $|B_i\rangle$ to $|A_{i+1}\rangle$, with hopping amplitude $\Omega_{B'}(t)$.  With this choice of rotating-frame phases, the resonant Hamiltonian becomes
\begin{equation}
\begin{aligned}
\hat H(t)
=&
\sum_{i=1}^{N}
\sum_{\mu=B,C,D}
\Omega_{\mu}(t)
|A_i\rangle\langle \mu_i|
\\
&+
\sum_{i=1}^{N-1}
\Omega_{B'}(t)
|A_{i+1}\rangle\langle B_i|
+
\mathrm{H.c.},
\end{aligned}
\label{eq:finite_lieb_hamiltonian_app}
\end{equation}
where $|\mu_i\rangle$ denotes $|B_i\rangle$, $|C_i\rangle$, or $|D_i\rangle$ for $\mu=B,C,D$.  Equation~\eqref{eq:finite_lieb_hamiltonian_app} is the finite Lieb Hamiltonian used in Eq.~\eqref{eq:finite_lieb_hamiltonian}.  Possible residual detunings, including Zeeman shifts or calibration errors, are represented by the diagonal terms $\Delta_r$ in Eq.~\eqref{eq:rydberg_rotating_hamiltonian}.

For reference, the periodic counterpart of the same Lieb connectivity is considered first.  In the Bloch basis $(|A_q\rangle,|B_q\rangle,|C_q\rangle,|D_q\rangle)$, the bulk Hamiltonian is
\begin{equation}
\hat H_{\rm bulk}(q,t)
=
\begin{pmatrix}
0 & g_B(q,t) & \Omega_C(t) & \Omega_D(t)\\
g_B^*(q,t) & 0 & 0 & 0\\
\Omega_C^*(t) & 0 & 0 & 0\\
\Omega_D^*(t) & 0 & 0 & 0
\end{pmatrix},
\end{equation}
where $g_B(q,t)=\Omega_B(t)+\Omega_{B'}(t)e^{-iq}$ for the chosen unit-cell convention. This bipartite form gives two zero-energy flat bands and two bright branches,
\begin{equation}
E_0^{(1,2)}(q,t)=0,
\qquad
E_\pm(q,t)=\pm \Omega_{\rm bulk}(q,t),
\end{equation}
where
$\Omega_{\rm bulk}(q,t)=\sqrt{|g_B(q,t)|^2+|\Omega_C(t)|^2+|\Omega_D(t)|^2}$.
The spectrum in Fig.~\ref{fig:fig1}(c) is this periodic reference.  It is used to identify the flat-band origin of the zero-energy sector; the pumping simulations themselves are performed with the finite open-boundary Hamiltonian in Eq.~\eqref{eq:finite_lieb_hamiltonian}.

The finite system inherits the same bipartite structure.  After ordering the Hilbert space as
$\mathcal H_A\oplus\mathcal H_{BCD}$, with
$\dim\mathcal H_A=N$ and $\dim\mathcal H_{BCD}=3N$, the resonant Hamiltonian takes the off-diagonal form
$\hat H=\begin{pmatrix}0&K\\K^\dagger&0\end{pmatrix}$,
where $K$ is an $N\times 3N$ coupling matrix from the $(B,C,D)$-sublattice to the $A$-sublattice.  For the nontrivial coupling configurations used in the pumping protocols, $K$ has full row rank, $\mathrm{rank}\,K=N$.  The zero-energy states are therefore supported on the larger $(B,C,D)$-sublattice and form a $2N$-dimensional null space of $K$.  For the three-cell lattice used in the main text, this gives six zero-energy states, as shown in Fig.~\ref{fig:fig1}(d).

Equivalently, writing the amplitudes on $|B_i\rangle$, $|C_i\rangle$, and $|D_i\rangle$ as $b_i$, $c_i$, and $d_i$, the zero-energy condition at each $|A_i\rangle$ imposes one linear constraint,
$\Omega_B b_i+\Omega_C c_i+\Omega_D d_i+\Omega_{B'}b_{i-1}=0$ with $b_0=0$.  Thus the $3N$ amplitudes on the $(B,C,D)$-sublattice are subject to $N$ independent constraints, leaving $2N$ independent zero-energy states.

\section{Local adiabatic reduction and global adiabatic criterion}
\label{app:local_three_level_reduction}
\label{app:gac_theory}
\subsection{Local $\Lambda$-type adiabatic passages}
The first transfer step of $P_1$ provides a representative local reduction. The relevant sector is $\{|C_2\rangle,|A_2\rangle,|B_1\rangle\}$, where the active microwave-induced couplings connect $|A_2\rangle$ with $|C_2\rangle$ through $\Omega_C(t)$ and $|A_2\rangle$ with $|B_1\rangle$ through $\Omega_{B'}(t)$. In the ordered basis $(|C_2\rangle,|A_2\rangle,|B_1\rangle)$, the local Hamiltonian is
\begin{equation}
\hat H_{\Lambda}(t)
=
\begin{pmatrix}
0 & \Omega_C(t) & 0\\
\Omega_C(t) & 0 & \Omega_{B'}(t)\\
0 & \Omega_{B'}(t) & 0
\end{pmatrix}.
\end{equation}
Here the couplings are taken real, as in the simulations; phases can be restored by the same rotating-frame convention used above.

The zero-energy state of this local Hamiltonian has no $|A_2\rangle$ component. Solving $\hat H_{\Lambda}(t)|d_{\Lambda}(t)\rangle=0$ gives
\begin{equation}
\begin{aligned}
|d_{\Lambda}(t)\rangle
&=
\frac{
\Omega_{B'}(t)|C_2\rangle
-
\Omega_C(t)|B_1\rangle
}{
\Omega_{\Lambda}(t)
},
\\
\Omega_{\Lambda}(t)
&=
\sqrt{
|\Omega_C(t)|^2+|\Omega_{B'}(t)|^2
}.
\end{aligned}
\end{equation}
For real couplings, this state may be parameterized by
\begin{equation}
\begin{aligned}
\tan\theta(t)
&=
\frac{\Omega_C(t)}{\Omega_{B'}(t)},
\\
|d_{\Lambda}(t)\rangle
&=
\cos\theta(t)|C_2\rangle
-
\sin\theta(t)|B_1\rangle .
\end{aligned}
\end{equation}
For this local three-level Hamiltonian, the two bright states are separated from the local zero-energy state by the energy scale $\Omega_\Lambda(t)$. The time dependence of the mixing angle induces nonadiabatic coupling from the local zero-energy state to the bright states. This local leakage tendency is characterized by
\begin{equation}
Q(t)
=
\frac{\sqrt{2}|\dot{\theta}(t)|}{\Omega_\Lambda(t)} .
\label{eq:Q_non}
\end{equation}
This is the nonadiabatic factor used in Sec.~\ref{sec:single_cycle} to construct timing indicators based on the GAC theory.
When $\Omega_{B'}$ precedes $\Omega_C$, the local zero-energy state evolves from $|C_2\rangle$ toward $-|B_1\rangle$, up to the overall phase convention. This local three-level reduction is used only to identify the transfer window and timing criterion. The single-cycle traces, composed-cycle dynamics, and robustness calculations in the main text are obtained from the complete twelve-level Hamiltonian, with the open-system loss model included where specified.

\subsection{Global adiabatic criterion}
The GAC used for the timing design in Fig.~\ref{fig:fig3} is summarized below.  The derivation is formulated for a general time-dependent Hamiltonian and is then applied in the main text to the representative local $\Lambda$-type transfer step.  The quantity $\varepsilon$ denotes the total nonadiabatic transition probability from the instantaneous target/dark state to the other instantaneous modes.

A conservative estimate for the total nonadiabatic transition probability is obtained as follows.  The starting point is the Schr\"odinger equation
\begin{equation}
i \partial_t |\Psi(t)\rangle = \hat H(t) |\Psi(t)\rangle .
\label{eq:gac_schrodinger}
\end{equation}
The state vector can be expanded in the instantaneous eigenbasis of $\hat H(t)$
\begin{equation}
|\Psi(t)\rangle
=
\sum_n a_n(t)|\psi_n(t)\rangle ,
\label{eq:gac_expansion}
\end{equation}
where $|\psi_0(t)\rangle\equiv|\psi_{\rm d}(t)\rangle$ is the instantaneous target/dark state and $|\psi_m(t)\rangle$ with $m\neq0$ are other bright states.  The corresponding eigenvalues are $E_0(t)$ and $E_m(t)$, and the instantaneous eigenstates satisfy $\langle\psi_m(t)|\psi_n(t)\rangle=\delta_{mn}$.

Projecting Eq.~\eqref{eq:gac_schrodinger} onto a bright state and retaining the leading contribution from the instantaneous target/dark state gives
\begin{equation}
\partial_t a_m(t)
\simeq
-i\Delta E_m(t)a_m(t)
-
\langle\psi_m(t)|\partial_t\psi_{\rm d}(t)\rangle ,
\label{eq:gac_amp_eq}
\end{equation}
where $\Delta E_m(t)=E_m(t)-E_0(t)$.  With the initial condition $a_m(0)=0$, the formal solution is
\begin{equation}
a_m(T)
=
-\int_0^T
\langle\psi_m(t)|\partial_t\psi_{\rm d}(t)\rangle
e^{i\theta_m(t)}
\,dt ,
\end{equation}
where $\theta_m(t)=\int_0^t\Delta E_m(t')\,dt'$ is the dynamical phase.  The total nonadiabatic transition probability to all bright states is
\begin{equation}
\varepsilon
=
\sum_{m\neq0}|a_m(T)|^2 .
\label{eq:gac_total_prob}
\end{equation}

Applying the triangle inequality removes the unit-modulus dynamical phase and bounds the transition amplitudes by the time-integrated nonadiabatic couplings.  Using $\sum_m|a_m|^2\le(\sum_m|a_m|)^2$, one obtains
\begin{equation}
\varepsilon
\le
\left(
\sum_{m\neq0}
\int_0^T
\left|
\langle\psi_m(t)|\partial_t\psi_{\rm d}(t)\rangle
\right|
\,dt
\right)^2 .
\end{equation}

The instantaneous nonadiabaticity factor is defined as
\begin{equation}
Q(t)
=
\sum_{m\neq0}
\left|
\frac{
\langle\psi_m(t)|\partial_t\psi_{\rm d}(t)\rangle
}{
\Delta E_m(t)
}
\right| ,
\label{eq:gac_q_def}
\end{equation}
which measures the total nonadiabatic coupling normalized by the instantaneous energy separations.  Taking a conservative gap-scale estimate in which the relevant energy scale is represented by the minimum gap
\(\Delta E_{\min}\equiv\min_{t,m}|\Delta E_m(t)|\), and in which the nonadiabatic contributions are assumed to add constructively, gives
\begin{equation}
\varepsilon
\le
\left(
\Delta E_{\min}
\int_0^T Q(t)\,dt
\right)^2 .
\label{eq:gac_bound}
\end{equation}

The global criterion follows by bounding the time integral of $Q(t)$.  The Cauchy--Schwarz inequality gives
\begin{equation}
\begin{aligned}
\left(
\int_0^T Q(t)\,dt
\right)^2
&\le
\int_0^T 1^2\,dt
\int_0^T Q(t)^2\,dt
\\
&=
T\int_0^T Q(t)^2\,dt .
\end{aligned}
\end{equation}
The mean and temporal variance of $Q(t)$ are defined as
\begin{equation}
\overline Q
=
\frac{1}{T}\int_0^T Q(t)\,dt,
\qquad
\sigma_Q^2
=
\frac{1}{T}
\int_0^T
\left[
Q(t)-\overline Q
\right]^2
\,dt .
\label{eq:gac_mean_var}
\end{equation}
By definition, the second moment satisfies $\int_0^T Q(t)^2\,dt=T(\overline Q^2+\sigma_Q^2)$.
Substituting this result into the Cauchy--Schwarz bound yields
\begin{equation}
\int_0^T Q(t)\,dt
\le
T\sqrt{\overline Q^2+\sigma_Q^2}.
\end{equation}
Combining this result with Eq.~\eqref{eq:gac_bound} gives the GAC estimate
\begin{equation}
\varepsilon
\le
\left[
\Delta E_{\min}
T
\sqrt{\overline Q^2+\sigma_Q^2}
\right]^2 .
\label{eq:gac_rms_bound}
\end{equation}

This estimate should therefore be regarded as a conservative design criterion rather than a tight prediction of leakage, since its derivation neglects possible phase cancellations.  Its practical value is to identify a global cost controlled by both the mean nonadiabatic burden $\overline Q$ and its temporal variance $\sigma_Q^2$, thereby penalizing schedules in which nonadiabaticity is concentrated in sharp temporal peaks.

\section{Open-system model and numerical implementation}
\label{app:open_system_master_equation}

The open-system simulations in Figs.~\ref{fig:fig4} and~\ref{fig:fig5} are performed in a 13-dimensional Hilbert space consisting of the twelve selected Rydberg working states and one auxiliary sink state $|\mathrm{loss}\rangle$.  The Hamiltonian is expressed in units of the peak microwave coupling $\Omega_0$, and time is expressed in units of $1/\Omega_0$.  For the physical conversion used in the figures, we take $\Omega_0/2\pi=2.0\,\mathrm{MHz}$~\cite{TChen2024PRL,Chen2024RydbergFlux,Chen2025RydbergCage}.

The state-dependent Rydberg lifetimes and transition parameters are estimated with the ARC for the selected ${}^{39}\mathrm{K}$ levels~\cite{Sibalic2017ARC}. The density matrix is evolved with the Lindblad master equation
\begin{equation}
\dot{\rho}(t)
=
-i[\hat H(t),\rho(t)]
+
\sum_j
\left[
L_j \rho(t) L_j^\dagger
-\frac{1}{2}
\left\{
L_j^\dagger L_j,\rho(t)
\right\}
\right],
\end{equation}
where $\hat H(t)$ is the time-dependent microwave-coupled Lieb-lattice Hamiltonian.  Radiative loss from each selected Rydberg state is modeled by
\begin{equation}
L_j
=
\sqrt{\gamma_j}
|\mathrm{loss}\rangle\langle j| ,
\end{equation}
where $|j\rangle$ denotes one of the twelve selected working states.  The decay rates $\gamma_j$ are obtained from the $10~\mathrm{K}$ lifetime estimates.  Across the selected states, the lifetimes span $\tau_{10K}=72.66\sim276.41\,\mu\mathrm{s}$, corresponding to the dimensionless decay-rate range $\gamma_{10K}=2.879\times10^{-4}\sim1.095\times10^{-3}$. The master equation is solved numerically using QuTiP~\cite{Lambert2026QuTiP5}.  The performance measure used in the open-system calculations is the target-state population $P_{\mathrm{tar}}=\rho_{\mathrm{work}}(T)_{j_{\mathrm{tar}},j_{\mathrm{tar}}}$, where $\rho_{\mathrm{work}}(T)$ is the final density matrix restricted to the twelve selected Rydberg working states, excluding the auxiliary loss state.  The target index $j_{\mathrm{tar}}$ is chosen according to the corresponding composed-cycle mapping for each input state and temporal order.  This quantity is evaluated without normalization by the survival probability, and therefore includes coherent transfer errors, leakage to bright states, and dissipative loss.

The perturbation scans in Fig.~\ref{fig:fig5} are implemented on top of the same open-system evolution.  A global amplitude error rescales all microwave-induced couplings as $\Omega_g(t)\mapsto(1+\epsilon_\Omega)\Omega_g(t)$, with $g\in\{B,C,D,B'\}$.  A deterministic detuning gradient is added as a diagonal offset $\Delta_j=\eta x_j$, where $x_j$ labels the unit-cell position of the selected state.  The deterministic scans use $\epsilon_\Omega\in[-0.20,0.20]$ and $\eta/\Omega_0\in[-0.20,0.20]$.

For static random on-site disorder, each realization is generated by sampling diagonal offsets uniformly from $[-W_{\mathrm{on}},W_{\mathrm{on}}]$, followed by subtracting the sample mean so that the disorder is zero mean within that realization.  For static coupling disorder, each physical coupling edge is multiplied by $1+\delta_e$, with $\delta_e$ sampled uniformly from $[-W_J,W_J]$.  The random-disorder scans use $W_{\mathrm{on}}/\Omega_0\in[0,0.20]$, $W_J\in[0,0.20]$, and $N_{\mathrm{real}}=200$ realizations.  For each disorder strength, $P_{\mathrm{tar}}$ is first averaged over the selected input-state and sequence-order cases and then averaged over disorder realizations.  The shaded regions in Figs.~\ref{fig:fig5}(c) and \ref{fig:fig5}(d) denote $\pm 1$ standard deviation over these disorder realizations after the input-sequence averaging.
\bibliography{ref}

@misc{wu2026globaladiabaticcriterionfast,
      title={Global adiabatic criterion for fast topological photon transfer in {F}ock-state lattices}, 
      author={Jin-Lei Wu and Pei-Yao Song and Jia Li and Ya Gao and Yan Wang and Shi-Lei Su},
      year={2026},
      eprint={2606.03409},
      archivePrefix={arXiv},
      primaryClass={quant-ph},
      url={https://arxiv.org/abs/2606.03409}, 
}

@article{Demirplak2003JPCA,
author = {Demirplak, Mustafa and Rice, Stuart A.},
title = {Adiabatic Population Transfer with Control Fields},
journal = {The Journal of Physical Chemistry A},
volume = {107},
number = {46},
pages = {9937-9945},
year = {2003},
doi = {10.1021/jp030708a},
URL = {https://doi.org/10.1021/jp030708a}
}

@article{TChen2024PRL,
  title = {Quantum Walks and Correlated Dynamics in an Interacting Synthetic Rydberg Lattice},
  author = {Chen, Tao and Huang, Chenxi and Gadway, Bryce and Covey, Jacob P.},
  journal = {Phys. Rev. Lett.},
  volume = {133},
  issue = {12},
  pages = {120604},
  numpages = {7},
  year = {2024},
  month = {Sep},
  publisher = {American Physical Society},
  doi = {10.1103/PhysRevLett.133.120604},
  url = {https://link.aps.org/doi/10.1103/PhysRevLett.133.120604}
}

@misc{huang2025interactionassistedtopologicalpumpingfew,
      title={Interaction-assisted topological pumping in few- and many-atom Rydberg arrays}, 
      author={Chenxi Huang and Tao Chen and Qian Liang and Matthew A. Krebs and Ethan Springhorn and Ruiyu Li and Mingsheng Tian and Kaden R. A. Hazzard and Jacob P. Covey and Bryce Gadway},
      year={2025},
      eprint={2512.12364},
      archivePrefix={arXiv},
      primaryClass={cond-mat.quant-gas},
      url={https://arxiv.org/abs/2512.12364}
}

@article{YLu2024PRA,
  title = {Wave-packet dynamics and long-range tunneling within the Su-Schrieffer-Heeger model using Rydberg-atom synthetic dimensions},
  author = {Lu, Y. and Wang, C. and Kanungo, S. K. and Yoshida, S. and Dunning, F. B. and Killian, T. C.},
  journal = {Phys. Rev. A},
  volume = {109},
  issue = {3},
  pages = {032801},
  numpages = {7},
  year = {2024},
  month = {Mar},
  publisher = {American Physical Society},
  doi = {10.1103/PhysRevA.109.032801},
  url = {https://link.aps.org/doi/10.1103/PhysRevA.109.032801}
}

@article{YLu2024PRA2,
  title = {Probing the topological phase transition in the Su-Schrieffer-Heeger Hamiltonian using Rydberg-atom synthetic dimensions},
  author = {Lu, Y. and Wang, C. and Kanungo, S. K. and Dunning, F. B. and Killian, T. C.},
  journal = {Phys. Rev. A},
  volume = {110},
  issue = {2},
  pages = {023318},
  numpages = {6},
  year = {2024},
  month = {Aug},
  publisher = {American Physical Society},
  doi = {10.1103/PhysRevA.110.023318},
  url = {https://link.aps.org/doi/10.1103/PhysRevA.110.023318}
}

@Article{YChen2025NC,
  author        = {Chen, Youlve and Fan, Yunru and Larsonneur, Gulliver and Xiang, Jinlong and He, An and Wang, Guohuai and Zhang, Xu-Lin and Ma, Guancong and Zhou, Qiang and Guo, Guangcan and Su, Yikai and Guo, Xuhan},
  title         = {High-dimensional non-{A}belian holonomy in integrated photonics},
  journal       = {Nat. Commun.},
  year          = {2025},
  volume        = {16},
  number        = {1},
  pages         = {3650},
  month         = apr,
  issn          = {2041-1723},
  refid         = {Chen2025},
  url           = {https://doi.org/10.1038/s41467-025-58794-3}
}

@article{DYLi2025PI,
author = {Danying Yu and Wange Song and Luojia Wang and Rohith Srikanth and Sashank Kaushik Sridhar and Tao Chen and Chenxi Huang and Guangzhen Li and Xin Qiao and Xiaoxiong Wu and Zhaohui Dong and Yanyan He and Meng Xiao and Xianfeng Chen and Avik Dutt and Bryce Gadway and Luqi Yuan},
journal = {Photon. Insights},
number = {11},
pages = {R06},
publisher = {},
title = {Comprehensive review on developments of synthetic dimensions},
volume = {4},
month = {2},
year = {2025},
url = {https://www.researching.cn/articles/OJ5c82e7b039803e4}
}

@Article{Ozawa2019NRP,
  author        = {Ozawa, Tomoki and Price, Hannah M.},
  title         = {Topological quantum matter in synthetic dimensions},
  journal       = {Nat. Rev. Phys.},
  year          = {2019},
  volume        = {1},
  number        = {5},
  pages         = {349--357},
  month         = may,
  issn          = {2522-5820},
  refid         = {Ozawa2019},
  url           = {https://doi.org/10.1038/s42254-019-0045-3},
}

@article{ZLShan2025SB,
title = {Integrated photonics based on non-{A}belian holonomy},
journal = {Sci. Bull.},
volume = {70},
number = {23},
pages = {3927-3930},
year = {2025},
issn = {2095-9273},
doi = {https://doi.org/10.1016/j.scib.2025.10.022},
url = {https://www.sciencedirect.com/science/article/pii/S2095927325010448},
author = {Zhong-Lei Shan and Xu-Lin Zhang}
}

@article{ZZhu2024Science,
author = {Zijie Zhu  and Marius G\"achter  and Anne-Sophie Walter  and Konrad Viebahn  and Tilman Esslinger },
title = {Reversal of quantized {H}all drifts at noninteracting and interacting topological boundaries},
journal = {Science},
volume = {384},
number = {6693},
pages = {317-320},
year = {2024},
doi = {10.1126/science.adg3848},
URL = {https://www.science.org/doi/abs/10.1126/science.adg3848}
}

@Article{ZCheng2025NC,
  author        = {Cheng, Zheyu and Yue, Sijie and Long, Yang and Xie, Wentao and Yu, Zixuan and Teo, Hau Tian and Zhao, Y. X. and Xue, Haoran and Zhang, Baile},
  title         = {Observation of returning {T}houless pumping},
  journal       = {Nat. Commun.},
  year          = {2025},
  volume        = {16},
  number        = {1},
  pages         = {9669},
  month         = nov,
  issn          = {2041-1723},
  refid         = {Cheng2025},
  url           = {https://doi.org/10.1038/s41467-025-64671-w}
}

@article{QMo2025PRL,
  title = {Demonstration of Returning {T}houless Pump in a {B}erry Dipole System},
  author = {Mo, Qingyang and Liang, Shanjun and Lan, Xiangke and Zhu, Jie and Zhang, Shuang},
  journal = {Phys. Rev. Lett.},
  volume = {135},
  issue = {20},
  pages = {206603},
  numpages = {6},
  year = {2025},
  month = {Nov},
  publisher = {American Physical Society},
  doi = {10.1103/52yh-mlfm},
  url = {https://link.aps.org/doi/10.1103/52yh-mlfm}
}

@Article{YKSun2024NC,
  author   = {Sun, Yi-Ke and Shan, Zhong-Lei and Tian, Zhen-Nan and Chen, Qi-Dai and Zhang, Xu-Lin},
  title    = {{Two-dimensional non-Abelian Thouless pump}},
  journal  = {Nat. Commun.},
  year     = {2024},
  volume   = {15},
  number   = {1},
  pages    = {9311},
  month    = oct,
  issn     = {2041-1723},
  refid    = {Sun2024},
  url      = {https://doi.org/10.1038/s41467-024-53741-0}
}

@article{Danieli2025AVSQS,
    author = {Danieli, Carlo and Brosco, Valentina and Pilozzi, Laura and Citro, Roberta},
    title = {{Non-Abelian Thouless pumping in a Rice–Mele ladder}},
    journal = {AVS Quantum Sci.},
    volume = {7},
    number = {2},
    pages = {022001},
    year = {2025},
    month = {05},
  issn = {2639-0213},
    doi = {10.1116/5.0245350},
    url = {https://doi.org/10.1116/5.0245350}
}

@article{Brosco2021PRA,
  title = {{Non-Abelian Thouless pumping in a photonic lattice}},
  author = {Brosco, Valentina and Pilozzi, Laura and Fazio, Rosario and Conti, Claudio},
  journal = {Phys. Rev. A},
  volume = {103},
  issue = {6},
  pages = {063518},
  numpages = {10},
  year = {2021},
  month = {Jun},
  publisher = {American Physical Society},
  doi = {10.1103/PhysRevA.103.063518},
  url = {https://link.aps.org/doi/10.1103/PhysRevA.103.063518}
}

@article{Kraus2012PRL,
  title = {Topological States and Adiabatic Pumping in Quasicrystals},
  author = {Kraus, Yaacov E. and Lahini, Yoav and Ringel, Zohar and Verbin, Mor and Zilberberg, Oded},
  journal = {Phys. Rev. Lett.},
  volume = {109},
  issue = {10},
  pages = {106402},
  numpages = {5},
  year = {2012},
  month = {Sep},
  publisher = {American Physical Society},
  doi = {10.1103/PhysRevLett.109.106402},
  url = {https://link.aps.org/doi/10.1103/PhysRevLett.109.106402}
}

@Article{Lohse2018Nature,
author={Lohse, Michael
and Schweizer, Christian
and Price, Hannah M.
and Zilberberg, Oded
and Bloch, Immanuel},
title={Exploring 4{D }quantum {H}all physics with a 2{D} topological charge pump},
journal={Nature},
year={2018},
month={Jan},
day={01},
volume={553},
number={7686},
pages={55-58},
issn={1476-4687},
doi={10.1038/nature25000},
url={https://doi.org/10.1038/nature25000}
}

@Article{Zilberberg2018Nature,
author={Zilberberg, Oded
and Huang, Sheng
and Guglielmon, Jonathan
and Wang, Mohan
and Chen, Kevin P.
and Kraus, Yaacov E.
and Rechtsman, Mikael C.},
title={Photonic topological boundary pumping as a probe of 4{D} quantum {H}all physics},
journal={Nature},
year={2018},
month={Jan},
day={01},
volume={553},
number={7686},
pages={59-62},
issn={1476-4687},
doi={10.1038/nature25011},
url={https://doi.org/10.1038/nature25011}
}

@Article{Jurgensen2021Nature,
author={J{\"u}rgensen, Marius
and Mukherjee, Sebabrata
and Rechtsman, Mikael C.},
title={Quantized nonlinear {T}houless pumping},
journal={Nature},
year={2021},
month={Aug},
day={01},
volume={596},
number={7870},
pages={63-67},
issn={1476-4687},
doi={10.1038/s41586-021-03688-9},
url={https://doi.org/10.1038/s41586-021-03688-9}
}

@Article{Citro2023NRP,
author={Citro, Roberta
and Aidelsburger, Monika},
title={Thouless pumping and topology},
journal={Nature Reviews Physics},
year={2023},
month={Feb},
day={01},
volume={5},
number={2},
pages={87-101},
issn={2522-5820},
doi={10.1038/s42254-022-00545-0},
url={https://doi.org/10.1038/s42254-022-00545-0}
}

@article{Thouless1983Pumping,
  author = {Thouless, D. J.},
  title = {Quantization of particle transport},
  journal = {Phys. Rev. B},
  volume = {27},
  pages = {6083--6087},
  year = {1983},
  doi = {10.1103/PhysRevB.27.6083}
}

@article{Berry1984Phase,
  author = {Berry, M. V.},
  title = {Quantal phase factors accompanying adiabatic changes},
  journal = {Proc. R. Soc. Lond. A},
  volume = {392},
  pages = {45--57},
  year = {1984},
  doi = {10.1098/rspa.1984.0023}
}

@article{Simon1983Holonomy,
  author = {Simon, B.},
  title = {Holonomy, the Quantum Adiabatic Theorem, and {Berry}'s Phase},
  journal = {Phys. Rev. Lett.},
  volume = {51},
  pages = {2167--2170},
  year = {1983},
  doi = {10.1103/PhysRevLett.51.2167}
}

@article{Wilczek1984Holonomy,
  author = {Wilczek, F. and Zee, A.},
  title = {Appearance of Gauge Structure in Simple Dynamical Systems},
  journal = {Phys. Rev. Lett.},
  volume = {52},
  pages = {2111--2114},
  year = {1984},
  doi = {10.1103/PhysRevLett.52.2111}
}

@article{Sun2022NATP,
  author = {Sun, Y.-K. and Zhang, X.-L. and Yu, F. and Tian, Z. and Chen, Q. and Sun, H.},
  title = {{Non}-{Abelian} {Thouless} pumping in photonic waveguides},
  journal = {Nat. Phys.},
  volume = {18},
  pages = {1080--1085},
  year = {2022},
  doi = {10.1038/s41567-022-01669-x}
}

@article{You2022NonAbelianPump,
  author = {You, Oubo and Liang, Shanjun and Xie, Biye and Gao, Wenlong and Ye, Weimin and Zhu, Jie and Zhang, Shuang},
  title = {Observation of {Non}-{Abelian} {Thouless} Pump},
  journal = {Phys. Rev. Lett.},
  volume = {128},
  number = {24},
  pages = {244302},
  year = {2022},
  doi = {10.1103/PhysRevLett.128.244302}
}

@article{Kanungo2022RydbergSyntheticDimension,
  author = {Kanungo, S. K. and Whalen, J. D. and Lu, Y. and Yuan, M. and Dasgupta, S. and Dunning, F. B. and Hazzard, K. R. A. and Killian, T. C.},
  title = {Realizing topological edge states with {Rydberg}-atom synthetic dimensions},
  journal = {Nat. Commun.},
  volume = {13},
  pages = {972},
  year = {2022},
  doi = {10.1038/s41467-022-28550-y}
}

@article{Chen2024RydbergFlux,
  author = {Chen, T. and Huang, C. and Velkovsky, I. and Hazzard, K. and Covey, J. and Gadway, B.},
  title = {Strongly interacting {Rydberg} atoms in synthetic dimensions with a magnetic flux},
  journal = {Nat. Commun.},
  volume = {15},
  pages = {2675},
  year = {2024},
  doi = {10.1038/s41467-024-46823-6}
}

@article{Chen2025RydbergCage,
  author = {Chen, Tao and Huang, Chenxi and Velkovsky, Ivan and Ozawa, Tomoki and Price, Hannah and Covey, Jacob P. and Gadway, Bryce},
  title = {Interaction-driven breakdown of {Aharonov}--{Bohm} caging in flat-band {Rydberg} lattices},
  journal = {Nat. Phys.},
  volume = {21},
  pages = {221--227},
  year = {2025},
  doi = {10.1038/s41567-024-02714-7}
}

@article{Trautmann2024RydbergThouless,
  author = {Trautmann, M. and Villadiego, I. S. and Deiglmayr, J.},
  title = {Realization of topological {Thouless} pumping in a synthetic {Rydberg} dimension},
  journal = {Phys. Rev. A},
  volume = {110},
  pages = {L040601},
  year = {2024},
  doi = {10.1103/PhysRevA.110.L040601},
  eprint = {2406.08551}
}

@article{Bergmann1998STIRAP,
  author = {Bergmann, K. and Theuer, H. and Shore, B. W.},
  title = {Coherent population transfer among quantum states of atoms and molecules},
  journal = {Rev. Mod. Phys.},
  volume = {70},
  pages = {1003--1025},
  year = {1998},
  doi = {10.1103/RevModPhys.70.1003}
}

@article{Rigolin2014DegenerateAdiabatic,
  author = {Rigolin, G. and Ortiz, G.},
  title = {Degenerate adiabatic perturbation theory: Foundations and applications},
  journal = {Phys. Rev. A},
  volume = {90},
  pages = {022104},
  year = {2014},
  doi = {10.1103/PhysRevA.90.022104}
}

@article{Beterov2009RydbergLifetime,
  author = {Beterov, I. I. and Ryabtsev, I. I. and Tretyakov, D. B. and Entin, V. M.},
  title = {Quasiclassical calculations of {BBR}-induced depopulation rates and effective lifetimes of {Rydberg} {nS}, {nP}, and {nD} alkali-metal atoms with n less than or equal to 80},
  journal = {Phys. Rev. A},
  volume = {79},
  pages = {052504},
  year = {2009},
  doi = {10.1103/PhysRevA.79.052504},
  eprint = {0810.0339}
}

@article{Leseleuc2018RydbergImperfections,
  author = {de Léséleuc, S. and Barredo, D. and Lienhard, V. and Browaeys, A. and Lahaye, T.},
  title = {Analysis of imperfections in the coherent optical excitation of single atoms to {Rydberg} states},
  journal = {Phys. Rev. A},
  volume = {97},
  pages = {053803},
  year = {2018},
  doi = {10.1103/PhysRevA.97.053803},
  eprint = {1802.10424}
}

@article{Walter2023HubbardThoulessBreakdown,
  author = {Walter, Anne-Sophie and Zhu, Zijie and G{\"a}chter, Marius and Minguzzi, Joaqu{\'i}n and Roschinski, Stephan and Sandholzer, Kilian and Viebahn, Konrad and Esslinger, Tilman},
  title = {Quantization and its breakdown in a {Hubbard}-{Thouless} pump},
  journal = {Nat. Phys.},
  volume = {19},
  pages = {1471--1475},
  year = {2023},
  doi = {10.1038/s41567-023-02145-w}
}

@article{Ostmann2018RydbergFlatBand,
  author = {Ostmann, Maike and Marcuzzi, Matteo and Minář, Jiří and Lesanovsky, Igor},
  title = {Synthetic lattices, flat bands and localization in {Rydberg} quantum simulators},
  journal = {Quantum Sci. Technol.},
  volume = {4},
  pages = {02LT01},
  year = {2019},
  doi = {10.1088/2058-9565/aaf29d}
}

@article{Scholl2022RydbergXXZ,
  author = {Scholl, P. and Williams, H. J. and Bornet, G. and Wallner, F. and Barredo, D. and Henriet, L. and Signoles, A. and Hainaut, C. and Franz, T. and Geier, S. and Tebben, A. and Salzinger, A. and Zürn, G. and Lahaye, T. and Weidemüller, M. and Browaeys, A.},
  title = {Microwave-engineering of programmable {XXZ} Hamiltonians in arrays of {Rydberg} atoms},
  journal = {PRX Quantum},
  volume = {3},
  pages = {020303},
  year = {2022},
  doi = {10.1103/PRXQuantum.3.020303}
}

@article{Wu2022RydbergGaugeFlux,
  author = {Wu, Xiaoling and Yang, Fan and Yang, Shuo and Mølmer, Klaus and Pohl, Thomas and Tey, Meng Khoon and You, Li},
  title = {Manipulating synthetic gauge fluxes via multicolor dressing of {Rydberg}-atom arrays},
  journal = {Phys. Rev. Research},
  volume = {4},
  pages = {L032046},
  year = {2022},
  doi = {10.1103/PhysRevResearch.4.L032046}
}

@article{Leroux2018NonAbelianAMO,
  author = {Leroux, Frédéric and Pandey, Kanhaiya and Rehbi, Riadh and Chevy, Frédéric and Miniatura, Christian and Grémaud, Benoît and Wilkowski, David},
  title = {{Non}-{Abelian} adiabatic geometric transformations in a cold strontium gas},
  journal = {Nat. Commun.},
  volume = {9},
  pages = {3580},
  year = {2018},
  doi = {10.1038/s41467-018-05865-3}
}

@article{Unanyan1999NonAbelianAdiabatic,
  author = {Unanyan, R. G. and Shore, B. W. and Bergmann, K.},
  title = {Laser-driven population transfer in four-level atoms: consequences of {Non}-{Abelian} geometrical adiabatic phase factors},
  journal = {Phys. Rev. A},
  volume = {59},
  pages = {2910--2919},
  year = {1999},
  doi = {10.1103/PhysRevA.59.2910}
}

@article{Chen2010ShortcutAdiabaticPassage,
  author = {Chen, Xi and Lizuain, I. and Ruschhaupt, A. and Guéry-Odelin, D. and Muga, J. G.},
  title = {Shortcut to adiabatic passage in two- and three-level atoms},
  journal = {Phys. Rev. Lett.},
  volume = {105},
  pages = {123003},
  year = {2010},
  doi = {10.1103/PhysRevLett.105.123003}
}

@article{GueryOdelin2019STA,
  author = {Guery-Odelin, D. and Ruschhaupt, A. and Kiely, A. and Torrontegui, E. and Martinez-Garaot, S. and Muga, J. G.},
  title = {Shortcuts to adiabaticity: Concepts, methods, and applications},
  journal = {Rev. Mod. Phys.},
  volume = {91},
  pages = {045001},
  year = {2019},
  doi = {10.1103/RevModPhys.91.045001}
}

@article{Hong2025MolecularRotationControl,
  author = {Hong, Qian-Qian and Dong, Daoyi and Henriksen, Niels E. and Nori, Franco and He, Jun and Shu, Chuan-Cun},
  title = {Precise quantum control of molecular rotation toward a desired orientation},
  journal = {Phys. Rev. Research},
  volume = {7},
  pages = {L012049},
  year = {2025},
  doi = {10.1103/PhysRevResearch.7.L012049}
}

@article{Zhou2026PTSSHQuantumBattery,
  author = {Zhou, A-Long and Xiao, Ya-Wen and Xu, Nuo and Gao, Li-Li and Li, Long-Jie and Zhou, Hang and Li, Zi-Min and Shu, Chuan-Cun},
  title = {Topological enhancement of a {$\mathcal{PT}$}-symmetric {Su}-{Schrieffer}-{Heeger} quantum battery},
  journal = {Phys. Rev. A},
  volume = {113},
  pages = {042213},
  year = {2026},
  doi = {10.1103/4klp-kw27}
}

@article{Fan2023MolecularPolaritonControl,
  author = {Fan, Li-Bao and Shu, Chuan-Cun and Dong, Daoyi and He, Jun and Henriksen, Niels E. and Nori, Franco},
  title = {Quantum Coherent Control of a Single Molecular-Polariton Rotation},
  journal = {Phys. Rev. Lett.},
  volume = {130},
  pages = {043604},
  year = {2023},
  doi = {10.1103/PhysRevLett.130.043604}
}

@article{Guo2024OptomechanicalSTA,
  author = {Guo, Jin-Kang and Wu, Jin-Lei and Cao, Ji and Zhang, Shou and Su, Shi-Lei},
  title = {Shortcut engineering for accelerating topological quantum state transfers in optomechanical lattices},
  journal = {Phys. Rev. A},
  volume = {110},
  pages = {043510},
  year = {2024},
  doi = {10.1103/PhysRevA.110.043510}
}

@article{Xiao2026GACPhotonicPumping,
  author = {Xiao, Kai-Heng and Su, Shi-Lei and Ni, Xiang and Sun, Yi-Ke and Guo, Jin-Kang and Hu, Zhi-Yong and Zhang, Xu-Lin and Li, Jia and Wu, Jin-Lei and Tian, Zhen-Nan and Chen, Qi-Dai},
  title = {Accelerated topological pumping in photonic waveguides based on global adiabatic criteria},
  journal = {arXiv preprint arXiv:2512.23466},
  year = {2026},
  doi = {10.48550/arXiv.2512.23466}
}

@article{Wu2025FloquetRydbergInteractions,
  author = {Wu, Jun and Wu, Jin-Lei and Guo, Fu-Qiang and Liu, Bing-Bing and Su, Shi-Lei and Song, Xue-Ke and Ye, Liu and Wang, Dong},
  title = {Quantum computation via {Floquet} tailored {Rydberg} interactions},
  journal = {npj Quantum Inf.},
  volume = {11},
  number = {1},
  pages = {118},
  year = {2025},
  doi = {10.1038/s41534-025-01068-z}
}

@article{Wu2020RydbergAmplitudeModulation,
  author = {Jin-Lei Wu and Shi-Lei Su and Yan Wang and Jie Song and Yan Xia and Yong-Yuan Jiang},
  journal = {Opt. Lett.},
  number = {5},
  pages = {1200--1203},
  publisher = {Optica Publishing Group},
  title = {Effective {Rabi} dynamics of {Rydberg} atoms and robust high-fidelity quantum gates with a resonant amplitude-modulation field},
  volume = {45},
  month = {Mar},
  year = {2020},
  url = {https://opg.optica.org/ol/abstract.cfm?URI=ol-45-5-1200},
  doi = {10.1364/OL.386765}
}

@article{Wu2021HolonomicEntanglingGates,
  author = {Wu, Jin-Lei and Wang, Yan and Han, Jin-Xuan and Jiang, Yongyuan and Song, Jie and Xia, Yan and Su, Shi-Lei and Li, Weibin},
  title = {Systematic-error-tolerant multiqubit holonomic entangling gates},
  journal = {Phys. Rev. Appl.},
  volume = {16},
  number = {6},
  pages = {064031},
  year = {2021},
  doi = {10.1103/PhysRevApplied.16.064031}
}

@article{Yun2024SiVGeometricGates,
  title = {Quantum computation in silicon-vacancy centers based on nonadiabatic geometric gates protected by dynamical decoupling},
  author = {Yun, M.-R. and Wu, Jin-Lei and Yan, L.-L. and Jia, Yu and Su, Shi-Lei and Shan, C.-X.},
  journal = {Phys. Rev. Appl.},
  volume = {21},
  issue = {6},
  pages = {064053},
  numpages = {18},
  year = {2024},
  month = {Jun},
  publisher = {American Physical Society},
  doi = {10.1103/PhysRevApplied.21.064053},
  url = {https://link.aps.org/doi/10.1103/PhysRevApplied.21.064053}
}

@article{Longhi2019PRB,
  title = {Topological pumping of edge states via adiabatic passage},
  author = {Longhi, Stefano},
  journal = {Phys. Rev. B},
  volume = {99},
  issue = {15},
  pages = {155150},
  numpages = {9},
  year = {2019},
  month = {Apr},
  publisher = {American Physical Society},
  doi = {10.1103/PhysRevB.99.155150},
  url = {https://link.aps.org/doi/10.1103/PhysRevB.99.155150}
}

@article{Tian2022PRL,
  title = {Experimental Realization of Nonreciprocal Adiabatic Transfer of Phonons in a Dynamically Modulated Nanomechanical Topological Insulator},
  author = {Tian, Tian and Zhang, Yichuan and Zhang, Liang and Wu, Longhao and Lin, Shaochun and Zhou, Jingwei and Duan, Chang-Kui and Jiang, Jian-Hua and Du, Jiangfeng},
  journal = {Phys. Rev. Lett.},
  volume = {129},
  issue = {21},
  pages = {215901},
  numpages = {6},
  year = {2022},
  month = {Nov},
  publisher = {American Physical Society},
  doi = {10.1103/PhysRevLett.129.215901},
  url = {https://link.aps.org/doi/10.1103/PhysRevLett.129.215901}
}

@article{Sibalic2017ARC,
  title = {{ARC}: An open-source library for calculating properties of alkali {Rydberg} atoms},
  author = {{\v{S}}ibali{\'c}, N. and Pritchard, J. D. and Adams, C. S. and Weatherill, K. J.},
  journal = {Computer Physics Communications},
  volume = {220},
  pages = {319--331},
  year = {2017},
  doi = {10.1016/j.cpc.2017.06.015}
}

@article{Lambert2026QuTiP5,
  title = {{QuTiP} 5: The Quantum Toolbox in {Python}},
  author = {Lambert, Neill and Gigu{\`e}re, Eric and Menczel, Paul and Li, Boxi and Hopf, Patrick and Su{\'a}rez, Gerardo and Gali, Marc and Lishman, Jake and Gadhvi, Rushiraj and Agarwal, Rochisha and Galicia, Asier and Shammah, Nathan and Nation, Paul and Johansson, J. R. and Ahmed, Shahnawaz and Cross, Simon and Pitchford, Alexander and Nori, Franco},
  journal = {Physics Reports},
  volume = {1153},
  pages = {1--62},
  year = {2026},
  doi = {10.1016/j.physrep.2025.10.001}
}

\end{document}